\documentclass[pdflatex,sn-mathphys-num]{sn-jnl}

\usepackage{graphicx}%
\usepackage{multirow}%
\usepackage{amsmath,amssymb,amsfonts}%
\usepackage{amsthm}%
\usepackage{mathrsfs}%
\usepackage[title]{appendix}%
\usepackage{xcolor}%
\usepackage{textcomp}%
\usepackage{manyfoot}%
\usepackage{booktabs}%
\usepackage{algorithm}%
\usepackage{algorithmicx}%
\usepackage{algpseudocode}%
\usepackage{listings}%
\usepackage{mathtools}

\DeclareMathOperator{\Tr}{Tr}

\theoremstyle{thmstyleone}%
\theoremstyle{thmstyletwo}%

\theoremstyle{thmstylethree}%

\begin{document}

\title[]{Influence of environment on quantum correlations in two-spin systems with dipole-dipole interactions}


\author*[1]{\fnm{G.A.} \sur{Bochkin}}\email{bochkin.g@yandex.ru}

\author[1]{\fnm{E.B.} \sur{Fel'dman}}\email{efeldman@icp.ac.ru}

\author[1]{\fnm{E.I.} \sur{Kuznetsova}}\email{kuznets@icp.ac.ru}

\author[1,2]{\fnm{E.I.} \sur{Shipulya}}\email{liza-shp@mail.ru}


\affil[1]{\orgdiv{Theoretical Department}, \orgname{Federal Research Center of Problems of Chemical Physics and Medicinal Chemistry of Russian Academy of Sciences}, \orgaddress{
\city{Chernogolovka}, \postcode{142432}, \state{Moscow Region}, \country{Russian Federation}}}

\affil[2]{ \orgname{Lomonosov Moscow State University}, \orgaddress{ \city{Moscow}, \postcode{119991}, \country{Russian Federation}}}



\abstract{An influence of environment on quantum correlations (entanglement and quantum discord) is studied in a two-spin-1/2 system with dipole-dipole interactions on the basis of Lindblad equation. We consider the simplest case when the environment causes only dephasing of system spins. The dependencies of entanglement and the quantum discord on the relaxation rate are obtained. We compare the influence of the environment on entanglement and quantum discord.}




\maketitle

\section{Introduction}\label{sec:intro}
Quantum correlations are the main resource of quantum computers which provide its advantages over their classical counterparts \cite{Nielsen}. Initiall, entanglement has been considered as a measure of quantum correlations \cite{amico,horodecki}. Quantum entanglement is able to bind parts of composite systems even when no interaction between those parts (the Einstein-Podolsky-Rosen paradox \cite{epr}). However, in 2001 Henderson and Vedral \cite{henderson} and independently Olivier and Zurek \cite{ollivier} showed that entanglement does not measure all quantum correlations in the system. They have found that all correlations in the system can be separated into a purely classical part and a purely quantum part which was named quantum discord. Soon, various remarkable properties of the quantum discord were discovered. For example it has been shown that quantum discord can be used to detect quantum phase transitions \cite{dillen_,sarandy}. An interesting fact is that almost all quantum states have non-classical correlations \cite{ferraro}. It has been shown also that the quantum discord is an important resource for designing secure communication channels \cite{madhok}. Published experimental data have confirmed that quantum benefits can be obtained using the quantum discord without entanglement \cite{dakich,gu2012}.

Calculation of entanglement in binary spin systems can be done using a formula \cite{hill, wootters} and many publications for such systems are given in \cite{horodecki,doronin07,furman,aldoshin}. Unfortunately, calculation of the quantum discord remains very difficult as it requires optimization over performed measurements \cite{luo,ali}. Among various systems used for calculation of entanglement and quantum discord we note dipolar systems \cite{furman,kuznets13} (systems with dipolar interactions \cite{abragam}), which we consider in the present article.

The primary aim of the present article is the investigation of the influence of the environment on the quantum correlations. In practice, the interaction of the particular system of interest (in the considered case of the two-spin system) with the environment is inevitable, and our task is to take this into account. Our approach is based on the Lindblad equation, which is commonly used in the theory of open quantum systems \cite{preskill,manzano}. The Lindblad equation is only applicable if stochastic interactions with the environment on relevant time scales can be described as a Markov process \cite{preskill}. The application of the Lindblad equation guarantees that the density matrix remains a valid density matrix, that is, Hermitian, positive semidefinite and with trace of one \cite{preskill}, even if it is converted to superoperator form, in contrast to the standard NMR relaxation methods \cite{abragam}.

We solve the Lindblad equation analytically for our two-spin system, continuing our work \cite{feldman_pan,bochkin24,feldman_ship} of investigation of relaxation in multiple quantum NMR experiments \cite{baum}.

The paper is organized as follows. In Sect.\ref{sec:analyt} we analytically solve the Lindblad  equation for the density matrix of our two-spin system. The system consists of two spins coupled by dipolar interaction \cite{abragam} subject to dephasing relaxation described by Lindblad operators \cite{preskill,manzano}. The obtained solution is used in Sect.\ref{sec:entangl} for our two-spin system. The dependencies of entanglement on the evolution time and the temperature are also investigated. The analogous investigation of the quantum discord is performed in Sect.\ref{sec:discord}. Brief discussion of our results is given in the concluding section \ref{sec:concl}.

\section{The analytical solution of the Lindblad equation for the dipolar two-spin system with dephasing relaxation}\label{sec:analyt}
We consider a system which consists of two spins ($s=1/2$) coupled by dipolar interaction. The Hamiltonian $H_{dz}$ of the system reads
\begin{equation}
H_{dz}=D(2I_{z1}I_{z2} - I_{1x}I_{2x} - I_{1y}I_{2y})\label{ham},
\end{equation}
where $D$ is the dipolar coupling constant and $I_{\alpha i}$ ($\alpha=x,y,z$; $i=1,2$) are the components of the spin angular momentum operator. The system is initially in the thermal equilibrium state, which can be written as
\begin{equation}
\rho=\frac{\exp(\beta I_z)}Z,\qquad Z=e^\beta+e^{-\beta}+2=4\cosh^2\frac\beta 2\label{rhoinittrue},
\end{equation}
where $I_z = I_{z1} +I_{z2}$, $Z$ is the partition function, and $\beta$ is a dimensionless parameter inversely proportional to the temperature. Following the NMR experimental conditions \cite{goldman} we apply $\pi/2$ RF pulse at time instant $t=0$. The state of the system after the pulse is
\begin{equation}
\rho(0)=\frac{\exp(\beta I_x)}Z,\qquad Z=4\cosh^2\frac\beta 2\label{rhoinit},
\end{equation}
where $I_x = I_{x1} +I_{x2}$. Below, we refer to this state as the initial state. For further calculations, we need the matrix representation of \eqref{rhoinit}:

\begin{equation}
\rho(0)=\frac{1}{4} \left(
\begin{array}{cccc}
 1 & \mathrm{th} \left(\frac{\beta }{2}\right) & \mathrm{th} \left(\frac{\beta }{2}\right) & \mathrm{th} ^2\left(\frac{\beta }{2}\right) \\
 \mathrm{th} \left(\frac{\beta }{2}\right) & 1 & \mathrm{th} ^2\left(\frac{\beta }{2}\right) & \mathrm{th} \left(\frac{\beta }{2}\right) \\
 \mathrm{th} \left(\frac{\beta }{2}\right) & \mathrm{th} ^2\left(\frac{\beta }{2}\right) & 1 & \mathrm{th} \left(\frac{\beta }{2}\right) \\
 \mathrm{th} ^2\left(\frac{\beta }{2}\right) & \mathrm{th} \left(\frac{\beta }{2}\right) & \mathrm{th} \left(\frac{\beta }{2}\right) & 1 \\
\end{array}
\right)\label{rhoinitmatrix}
\end{equation}
During the evolution of the system, it also interacts with the environment. We consider a case in which this interaction only causes dephasing of the system spins. Then the overall evolution of the system, including relaxation, can be described by the Lindblad equation \cite{preskill,manzano}:
\begin{equation}
\frac{d\rho}{dt}=-i\left[H_{dz},\rho(t)\right]+g\sum\limits_{k=1}^2 \left\{I_{zk}\rho(t)I_{zk}- \frac 12 I_{zk}^2\rho(t)-\frac 12\rho(t)I_{zk}^2 \right\}\label{lindbl}
\end{equation}
where $g$ is a constant, which characterizes the relaxation rate and has the dimension of $t^{-1}$. Below, we will use the dimensionless time $Dt$ and the dimensionless relaxation rate $g/D$. The matrix form of Eq.\eqref{lindbl} reads \small 
\begin{equation}
\frac{d\rho}{dt}=
\begingroup
\left(
\begin{array}{cccc}
 0 & \begin{multlined}-\frac{1}{2} (g+2 i) \rho_{12}(t)-\\ \frac{1}{2} i \rho_{13}(t)\end{multlined} & \begin{multlined}-\frac{1}{2} i \rho_{12}(t) -\\ \left(i+\frac{g}{2}\right) \rho_{13}(t)\end{multlined} & -g \rho_{14}(t) \\[40pt]
 \begin{multlined}\frac{1}{2} i \rho_{31}(t) -\\ \frac{(g-2 i)}{2} \rho_{21}(t))\end{multlined} &\begin{multlined} -\frac{1}{2} i (\rho_{23}(t)-\rho_{32}(t))\end{multlined} & \begin{multlined} - g \rho_{23}(t) -\\ \frac{1}{2} i(\rho_{22}(t)-\rho_{33}(t))\end{multlined} & \begin{multlined}\frac{i}{2}  \rho_{34}(t) +\\ \left(i-\frac{g}{2}\right) \rho_{24}(t)\end{multlined} \\[40pt]
 \begin{multlined}\frac{1}{2} i \rho_{21}(t)+\\ \left(i-\frac{g}{2}\right) \rho_{31}(t)\end{multlined} & \begin{multlined}- g \rho_{32}(t)+\\ \frac{1}{2} i (\rho_{22}(t)-\rho_{33}(t))\end{multlined} & \begin{multlined}\frac{1}{2} i (\rho_{23}(t)-\rho_{32}(t))\end{multlined} & \begin{multlined}\frac{i}{2}  \rho_{24}(t)+\\ \left(i-\frac{g}{2} \right) \rho_{34}(t)\end{multlined} \\[40pt]
 -g \rho_{41}(t) & \begin{multlined}-\left(\frac{g}{2}+i \right) \rho_{42}(t)-\\ \frac{1}{2} i \rho_{43}(t)\end{multlined} & \begin{multlined}-\frac{1}{2} i \rho_{42}(t)-\\\left(i+\frac{ g}{2}\right) \rho_{43}(t))\end{multlined} & 0 
\end{array}
\right)
\endgroup
\label{lindbl_matrix}
\end{equation} \normalsize
The equation \eqref{lindbl_matrix} can be separated into subsystems. There are four subsystems of two equations each:

\begin{align}\frac d{dt}\left(
\begin{array}{c}
 \rho_{42} \\
 \rho_{43} \\
\end{array}
\right)=\left(
\begin{array}{cc}
 -\frac{g}{2}-i & -\frac{i}{2} \\
 -\frac{i}{2} & -\frac{g}{2}-i \\
\end{array}
\right)\left(
\begin{array}{c}
 \rho_{42} \\
 \rho_{43} \\
\end{array}
\right),\,\frac d{dt}\left(
\begin{array}{c}
 \rho_{24} \\
 \rho_{34} \\
\end{array}
\right)=\left(
\begin{array}{cc}
 -\frac{g}{2}+i & \frac{i}{2} \\
 \frac{i}{2} & -\frac{g}{2}+i \\
\end{array}
\right)\left(
\begin{array}{c}
 \rho_{24} \\
 \rho_{34} \\
\end{array}
\right), \nonumber \\
\frac d{dt}\left(
\begin{array}{c}
 \rho_{21} \\
 \rho_{31} \\
\end{array}
\right)=\left(
\begin{array}{cc}
 -\frac{g}{2}+i & \frac{i}{2} \\
 \frac{i}{2} & -\frac{g}{2}+i \\
\end{array}
\right)\left(
\begin{array}{c}
 \rho_{21} \\
 \rho_{31} \\
\end{array}
\right),\,\frac d{dt}\left(
\begin{array}{c}
 \rho_{12} \\
 \rho_{13} \\
\end{array}
\right)=\left(
\begin{array}{cc}
 -\frac{g}{2}-i & -\frac{i}{2} \\
 -\frac{i}{2} & -\frac{g}{2}-i \\
\end{array}
\right)\left(
\begin{array}{c}
 \rho_{12} \\
 \rho_{13} \\
\end{array}
\right),\label{eqpairs}\end{align}
four separate equations
\begin{equation}\frac {d\rho_{41}}{dt}=-g \rho_{41}, \, \frac {d\rho_{14}}{dt} = -g\rho_{14}, \, \frac {d\rho_{11}}{dt} =\frac {d\rho_{44}}{dt} =0 \label{sep_eq}
\end{equation}
and one subsystem of four equations:
\begin{equation}\frac d{dt}\left(
\begin{array}{c}
 \rho_{22} \\
 \rho_{23} \\
 \rho_{32} \\
 \rho_{33} \\
\end{array}
\right)=\left(
\begin{array}{cccc}
 0 & -\frac{i}{2} & \frac{i}{2} & 0 \\
 -\frac{i}{2} & -g & 0 & \frac{i}{2} \\
 \frac{i}{2} & 0 & -g & -\frac{i}{2} \\
 0 & \frac{i}{2} & -\frac{i}{2} & 0 \\
\end{array}
\right)\left(
\begin{array}{c}
 \rho_{22} \\
 \rho_{23} \\
 \rho_{32} \\
 \rho_{33} \\
\end{array}
\right)\label{foureq}.\end{equation}
The solutions of \eqref{sep_eq} with the initial conditions \eqref{rhoinitmatrix} are obvious. Eqs.(\ref{eqpairs}) with the same initial conditions can be easily solved by change of variables to their sums and differences. In the case of the subsystem \eqref{foureq} we do the same, as follows:
\begin{equation}
\frac d{dt}\left(
\begin{array}{c}
 \rho_{22}+ \rho_{33} \\
 \rho_{23}+ \rho_{32} \\
 \rho_{23}-\rho_{32} \\
 \rho_{22}- \rho_{33} \\
\end{array}
\right)=\left(
\begin{array}{cccc}
 0 & 0 & 0 & 0 \\
 0 & -g & 0 & 0 \\
 0 & 0 & -g & -i \\
 0 & 0 & -i & 0 \\
\end{array}
\right)\left(
\begin{array}{c}
 \rho_{22}+ \rho_{33} \\
 \rho_{23}+ \rho_{32} \\
 \rho_{23}-\rho_{32} \\
 \rho_{22}- \rho_{33} \\
\end{array}
\right)\label{eqsumdif}
\end{equation}
One can find from Eqs.\eqref{eqsumdif},\eqref{rhoinitmatrix} that
\begin{equation}
\rho_{22}+ \rho_{33}=\frac 12, \, \rho_{23}+ \rho_{32}=2 \tanh^2\frac\beta 2 e^{-gt}, \,\rho_{22}- \rho_{33}=0, \, \rho_{23}-\rho_{32} = 0
\end{equation}
As a result, the solution of Eq.\eqref{lindbl_matrix} can be written as
\begin{equation}
\rho(t)=\left(
\begin{array}{cccc}
 \frac{1}{4} & \frac{1}{4} \tanh \left(\frac{\beta }{2}\right) e^{-\frac{1}{2} (g+3 i) t} & \frac{1}{4} \tanh \left(\frac{\beta }{2}\right) e^{-\frac{1}{2} (g+3 i) t} & \frac{1}{4} \tanh ^2\left(\frac{\beta }{2}\right) e^{-g t} \\
 \frac{1}{4} \tanh \left(\frac{\beta }{2}\right) e^{-\frac{1}{2} (g-3 i) t} & \frac{1}{4} & \frac{1}{4} \tanh ^2\left(\frac{\beta }{2}\right) e^{-g t} & \frac{1}{4} \tanh \left(\frac{\beta }{2}\right) e^{-\frac{1}{2} (g-3 i) t} \\
 \frac{1}{4} \tanh \left(\frac{\beta }{2}\right) e^{-\frac{1}{2} (g-3 i) t} & \frac{1}{4} \tanh ^2\left(\frac{\beta }{2}\right) e^{-g t} & \frac{1}{4} & \frac{1}{4} \tanh \left(\frac{\beta }{2}\right) e^{-\frac{1}{2} (g-3 i) t} \\
 \frac{1}{4} \tanh ^2\left(\frac{\beta }{2}\right) e^{-g t} & \frac{1}{4} \tanh \left(\frac{\beta }{2}\right) e^{-\frac{1}{2} (g+3 i) t} & \frac{1}{4} \tanh \left(\frac{\beta }{2}\right) e^{-\frac{1}{2} (g+3 i) t} & \frac{1}{4} \\
\end{array}
\right)\label{rhosol}
\end{equation}
It is obvious that this density matrix is centrosymmetric ($\rho_{i,j}=\rho_{5-i,5-j}$, $i,j= 1,2,3,4$) \cite{weaver,feldman12kuz}. That means the matrix can be simplified with the Hadamard transform $H_{ad}$ \cite{Nielsen}
\begin{equation}
H_{ad}=\frac 12\left(\begin{array}{cc}1 & 1 \\ 1& -1 \end{array}\right)\otimes \left(\begin{array}{cc}1 & 1 \\ 1& -1 \end{array}\right)
\end{equation}
and presented in the following form:
\begin{equation}
\rho_X=\left(
\begin{array}{cccc}
 a&0&0& \alpha \\ 0&b&0&0 \\ 0&0&c&0 \\ \alpha^* &0&0&d,
\end{array}
\right)\label{rhoxform}
\end{equation}
\begin{align}
a&= \frac{1}{4} \left(\tanh ^2\left(\frac{\beta }{2}\right) e^{-g t}+2\cos\frac{3t}2\tanh \left(\frac{\beta }{2}\right) e^{-\frac{1}{2} g t}+1\right),\nonumber \\
\alpha &= \frac{i}{2} \sin\frac{3t}2 \tanh \left(\frac{\beta }{2}\right) e^{-\frac{1}{2} (g+3 i) t}, \nonumber \\
b&= 
c= \frac{1}{4}-\frac{1}{4} \tanh ^2\left(\frac{\beta }{2}\right) e^{-g t}\label{rhox}
\end{align}
Eq.\eqref{rhoxform} shows that in another basis, the density matrix is an X matrix \cite{Nielsen}. It is important for the calculations of entanglement and quantum discord in the following sections.

\section{Entanglement in the dipolar two-spin system with dephasing relaxation}\label{sec:entangl}
We need to calculate the entanglement for the density matrix \eqref{rhoxform}. There is a useful auxiliary quantity, concurrence, which simplifies the calculation of entanglement \cite{hill, wootters} in two-spin systems.

\begin{figure}
\caption{Concurrence vs. the evolution time for various rates of dephasing relaxation}
\includegraphics{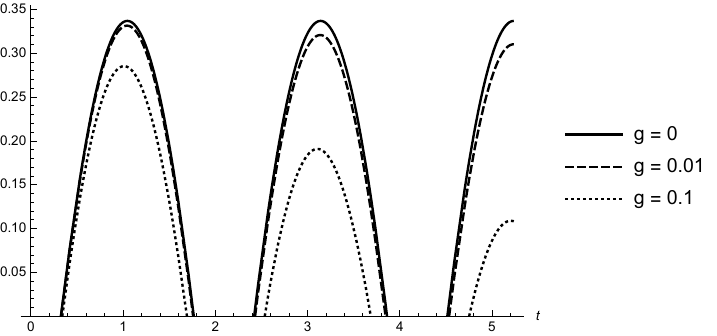}\label{fig:1}
\end{figure}

In order to determine the concurrence \cite{hill, wootters}, which describes entanglement, for the state described by the density matrix \eqref{rhoxform}, the eigenvalues of the following operator $R$ need to be found:
\begin{equation}
R=\rho_X(t)(\sigma_{y}\otimes \sigma_{y})\rho_X^*(t) (\sigma_{y}\otimes \sigma_{y})\label{sigmaprod},
\end{equation}
where $\sigma_{y}$ is the Pauli matrix. The matrix $\rho_X^*(t)$ is the complex conjugate of $\rho_X(t)$. The eigenvalues of the operator $R$ are 
\begin{multline}
\lambda_{1,2}= \\
\frac{1}{16} e^{-2 g t} \left(\pm 4 \tanh^2\frac\beta 2 e^{\frac{g t}{2}} \left| \sin \left(\frac{3 t}{2}\right)\right|  \sqrt{-2 \tanh^2\frac\beta 2 e^{g t} \cos (3 t)+e^{2 g t}+\tanh^4\frac\beta 2}+ \right . \\ \left .
\left(e^{g t}+\tanh^2\frac\beta 2\right)^2- 4 \tanh^2\frac\beta 2 e^{g t} \cos (3 t)\right) 
\end{multline}
\begin{equation}
\lambda_3=\lambda_4=\frac{1}{16} e^{-2 g t} \left(e^{g t}-\tanh^2\frac\beta 2\right)^2
\end{equation}
Obviously, $\lambda_1 \geq \lambda_2$. The inequality 
\begin{equation}
\left(e^{g t}+\tanh^2\frac\beta 2\right)^2 -4 \tanh^2\frac\beta 2 e^{g t} \cos 3t \geq \left(e^{g t}-\tanh^2\frac\beta 2\right)^2
\end{equation}
leads to the conclusion that $\lambda_1 \geq \lambda_3$. Therefore, $\lambda_1$ is the maximal eigenvalue of $R$. The concurrence can be found as  \cite{hill,wootters}
\begin{equation}
C=\max\left(0, \sqrt{\lambda_1} -\sqrt{\lambda_2}-\sqrt{\lambda_3}-\sqrt{\lambda_4}\right)\label{cformula}.
\end{equation}
Figure \ref{fig:1}
 shows the dependence of the concurrence on the dimensionless evolution time for various rates of dephasing relaxation and $\beta=1.5$.

\begin{figure}
\caption{Temperature dependence of maximum entanglement reached during evolution for various values of relaxation rate $g$}
\includegraphics{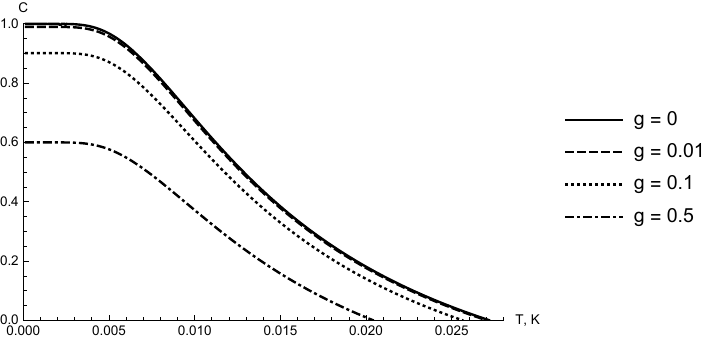}\label{fig:2}
\end{figure}

When there is no relaxation ($g=0$), concurrence is periodic. The dephasing relaxation disrupts the periodicity. The concurrence decreases as dephasing decreases the quantum correlations.
For obtaining the temperature dependence of the concurrence, we use the parameter $\beta=\frac{\hbar \omega_0}{kT}$, where $\hbar$ and $k$ are the reduced Planck's and the Boltzmann constants respectively, and $\omega_0$ is the angular Larmor frequency (we take $\omega_0 = 2\pi \cdot 500$~MHz, typical for modern NMR spectrometers). Figure~\ref{fig:2}
 shows the temperature dependence of the concurrence. One can see that entanglement appears when the temperature $T< 27$~mK. At higher temperatures, there is no entanglement.

\section{The quantum discord in the dipolar two-spin system with dephasing relaxation}\label{sec:discord}
Quantum discord, a measure of quantum correlation beyond entanglement, is the difference between total correlations and classical correlations between two subsystems (A and B) comprising the overall system \cite{ollivier, huang}. We designate the first spin as subsystem A, and the second spin as subsystem B, therefore we use notations $\rho_{AB}=\rho_X$, $\rho_A=Tr_B \rho_X$, $\rho_B=Tr_A \rho_X$. Classical correlation is defined as $J_B(\rho_{AB})=\max\limits_{\{\Pi_i\}} J_{\{\Pi_i\}}(\rho_{AB})$, where $J_{\{\Pi_i\}}(\rho_{AB}) = S(\rho_A)-\sum\limits_i p_i S(\rho_A^i)$ \cite{henderson}, $\{\Pi_i\}$ is the set of projectors defining a measurement (all measurements are performed on the subsystem B) , $p_i$ is the probability of measurement outcome $i$, $\rho_A^i=Tr_B(\Pi_i\rho_{AB})/p_i$ is the post-measurement state (for outcome $i$), $\rho'_{AB}=\sum\limits_i p_i\rho_A^i\otimes \Pi_i$ is the average post-measurement state of the overall system, $S$ is the von Neumann entropy ($S(\rho)=-\Tr \rho \log \rho$).
The maximization is performed over all possible von Neumann measurements. The discord is defined as \cite{ollivier, huang}
\begin{equation}
D_B(\rho_{AB})=\min\limits_{\{\Pi_i\}} S_B(\rho'_{AB})-S_B(\rho_{AB}),
\end{equation}
where $S_B(\rho_{AB})=S(\rho_{AB})-S(\rho_B)$ is the quantum conditional entropy.

\begin{figure}
\caption{Quantum discord for various values of dephasing relaxation rate $g$ and $\beta=1.5$.}
\includegraphics{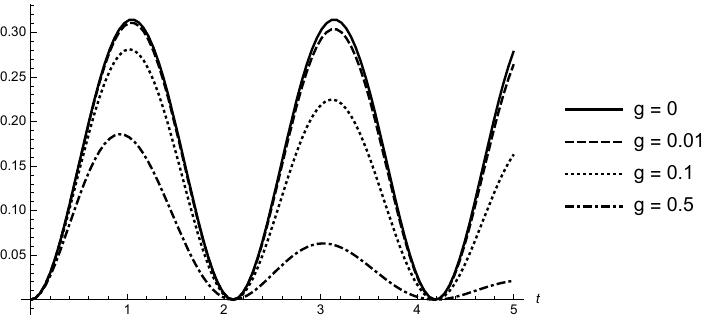}\label{fig:3}
\end{figure}

\begin{figure}
\caption{Maximum quantum discord vs. temperature for various values of dephasing relaxation rate $g$ and $\beta=1.5$.}
\includegraphics{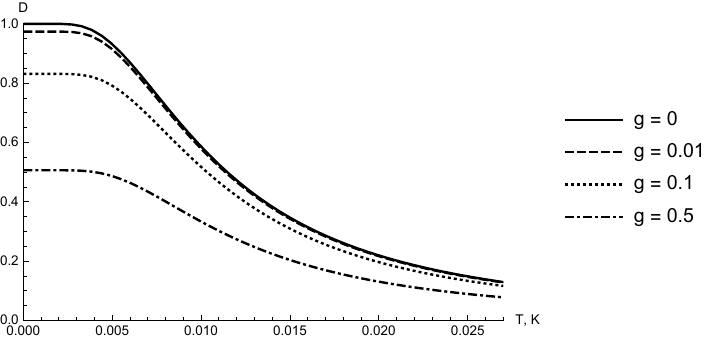}\label{fig:4}
\end{figure}

Further calculations follow \cite{huang}. Our density matrix \eqref{rhoxform} with elements \eqref{rhox} has the form considered in \cite{huang}, but additionally Huang's $\beta=0$ (not to be confused with our dimensionless temperature). As the result, we need only a single-variable optimization. We verified the analytical results numerically and obtained agreement within expected numerical error. The dependence of the quantum discord on time for several relaxation rates is shown in Fig.~\ref{fig:3}
, and Fig.~\ref{fig:4} shows the dependence on the temperature for various rates of dephasing relaxation.
Unlike concurrence, the discord is zero only for the infinite temperature or specific time instants, but otherwise the dependence is qualitatively similar: the discord decreases faster with increasing relaxation rate.
\section{Conclusions} \label{sec:concl}
Quantum correlations (entanglement and the quantum discord) were extensively investigated for a long time because they are important resources for quantum devices and are responsible for their advantages over classical counterparts. In the present work we studied the impact of the environment on quantum correlations.

We applied the Lindband equation for solving our problem and developed simple analytical method for their solution. Our approach is restricted to two-qubit systems. However, some of the developed methods could be applied to many-spin systems. It is very important to use the thermodynamic equilibrium state as the inital condition for the Lindblad equation. 

We studied the decay of entanglement and quantum discord due to dephasing relaxation and investigated the temperature dependence of the quantum correlations. The entanglement has a critical temperature above which is not present, but the discord is almost always present in our system, unlike the entanglement.

The developed approach opens new possibilities for investigation of quantum correlations in open systems.

\backmatter






\section*{Statements and Declarations}
\subsection*{Funding}
The work was performed as part of the state task, state registration no.~\mbox{124013000760-0}.
\subsection*{Competing Interests}
The authors have no relevant financial or non-financial interests to disclose.
\end{document}